# From Classical Hamiltonian Flow to Quantum Theory: Derivation of the Schrödinger Equation

Gerhard Grössing,
*Austrian Institute for Nonlinear Studies*,
Parkgasse 9, A-1030 Vienna, Austria

**Abstract:** It is shown how the essentials of quantum theory, i.e., the Schrödinger equation and the Heisenberg uncertainty relations, can be derived from classical physics. Next to the empirically grounded quantisation of energy and momentum, the only input is given by the assumption of fluctuations in energy and momentum to be added to the classical motion. Extending into the relativistic regime for spinless particles, this procedure leads also to a derivation of the Klein-Gordon equation. Comparing classical Hamiltonian flow with quantum theory, then, the essential difference is given by a vanishing divergence of the velocity of the probability current in the former, whereas the latter results from a much less stringent requirement, i.e., that only the average over fluctuations and positions of the average divergence be identical to zero.

## 1. Introduction

Classical mechanics can be considered as the geometrical-optical limiting case of a wave movement [1]: light (or other) rays orthogonal to wave fronts correspond to particle trajectories orthogonal to surfaces with constant "action function" $S$, where

$$S(\mathbf{x},\mathbf{p},t) = W(\mathbf{x},\mathbf{p}) - Et, \qquad (1.1)$$

with $\mathbf{x}, \mathbf{p}, t$ denoting location, momentum, and time coordinates, respectively, $E$ the energy, and $W$ the time-independent so-called "characteristic function". (Fig. 1)

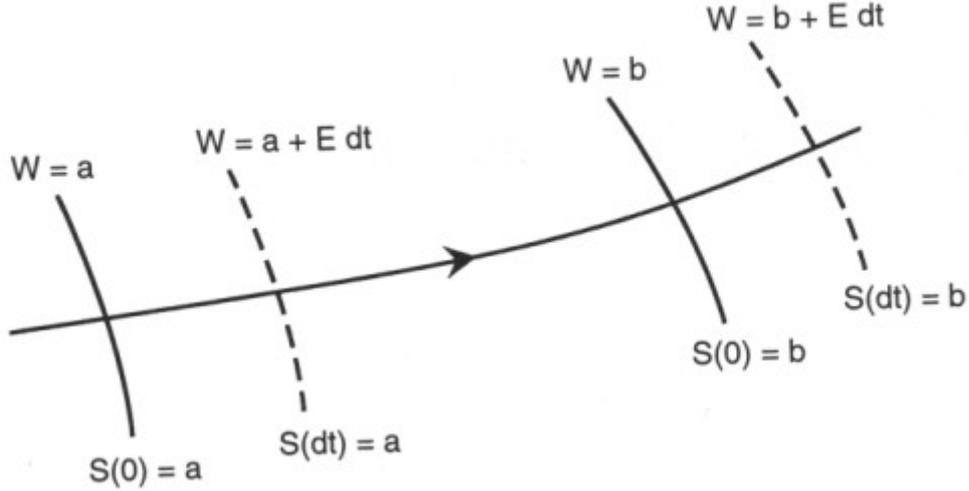

Fig. 1: Surfaces of constant action function $S$ representing wave fronts, with orthogonal particle trajectory.

The example of Fig. 1 is but a particular one. In accordance with Huygens' principle, another wide-spread example must be given by spherical wave surfaces. Here, the surface is initially concentrated at a point and then expands in a series of closed surfaces, such that the motion can be compared to that of a shock wave emanating from a "disturbing" point of a surface, i.e., as a travelling wavefront. (Fig. 2)

The velocity $u$ of the wave fronts can be calculated from the constancy of $S$, i.e.,

$$dS = \nabla S \cdot d\mathbf{x} + \left(\frac{\partial S}{\partial t}\right)dt = |\nabla S|\mathbf{n} \cdot d\mathbf{x} + \left(\frac{\partial S}{\partial t}\right)dt = 0, \qquad (1.2)$$

where

$$\mathbf{n} = \nabla S / |\nabla S| \qquad (1.3)$$

is a unit vector perpendicular to the surface $S = \text{constant}$ at point $\mathbf{x}$. As $\mathbf{n} \cdot d\mathbf{x} = ds$, the component of $d\mathbf{x}$ along the surface-normal, at time $t$ the speed of the wave-front at point $\mathbf{x}$ is given by

$$u(\mathbf{x},t) = \frac{ds}{dt} = -(\partial S/\partial t)/|\nabla S|. \qquad (1.4)$$

Note that $u$ does not coincide with the particle velocity, but can even be much larger than the vacuum speed of light $c$ as, for instance, when $(-\partial S/\partial t) = E = mc^2$ and $\nabla S = \mathbf{p} = m\mathbf{v}$. In principle, then, this could indicate a possible bridge towards quantum theory, whose nonlocal features might be brought into accordance with an (at least apparently) superluminal "spreading of information". Still, in classical mechanics, said particle trajectories (or "rays") and the propagating wave-fronts are not causally related to each other, and the wave fronts are seen mostly as a curious formal feature only.

Let us now turn to quantum theory and take as a starting point the above mentioned orthogonality in classical mechanics between wave fronts and particle trajectories. (For certain types of external potentials, this orthogonality is disturbed, but can be re-obtained with the introduction of a suitable metric. [2]) Erwin Schrödinger, in a talk given in 1952, has pointed out where to look for a transitional domain between classical wave mechanics and quantum mechanics:

> "At each place in a regularly propagating wave-train, one finds a two-fold structural connection of actions, which one may distinguish as 'longitudinal' and 'crosswise', respectively. The crosswise structure is the one along wave-planes; it manifests itself in diffraction and interference experiments. The longitudinal structure, in turn, is the one of the wave-normal and manifests itself when individual particles are observed. Both structural connections are completely confirmed, namely through useful experimental arrangements, each carefully designed for its particular purpose.
>
> However, the notions of 'longitudinal' and 'crosswise' structures are no precise ones, no absolute ones, just because the notions of wave-planes and wave-normals are not. These notions of wave-planes and wave-normals, and thus the differentiation between longitudinal and crosswise structures, are necessarily getting lost when the whole wave phenomenon is reduced to a small spatial region of the size of a single or only a few wavelengths. And this case now is of particular interest." [3]

However, we shall see shortly that said differentiation is not completely getting lost, essentially because a "wave phenomenon" in general cannot be reduced to a small region just by considering the main bulk of a wave packet only. In fact, it has been demonstrated conclusively in so-called "quantum post-selection experiments" [4, 5] that the handling of small-sized "wave packets" (to which Schrödinger's comment alludes) does not exclude the possibility to talk about wave-planes and particle trajectories: Interference between two possible quantum paths can be demonstrated, although the two corresponding bulks of the wave packets may never overlap (i.e., in $\mathbf{x}$-space). Rather, it is the non-locally extending plane-wave components of the wave-packets (which were often thought of as negligible), which make interference possible (i.e., in $\mathbf{k}$-space). In general, of course, the relation between wave-planes and particle paths will be more complicated than the simple orthogonality in classical mechanics.

Summarizing, one can say that the transition from classical to quantum mechanics is characterized by a transition from a general orthogonality between wave-fronts and particle trajectories, respectively, to a situation where orthogonality is an exception rather than the rule, the latter being a non-orthogonal relationship between particle trajectories and waves. In the next chapter, we shall consider a derivation of the Schrödinger (and the Klein-Gordon) equation under exactly this premise, i.e., thereby starting off with some basic equations of classical mechanics.

## 2. Derivation of the Schrödinger and Klein-Gordon equations from classical mechanics

Against wide-spread belief, it is possible to understand the foundations of quantum theory. In the following, I want to take a look back onto the early twentieth century and delineate a then possible line of developments towards the establishment of quantum theory, which is different from the line that actually did turn out. In doing so, however, we shall see that even nowadays one could have a very different, but still also very useful view of the quantum world.

Our working hypothesis is that the essentials of quantum theory can be derived from a) the empirical input of quantized energy and momentum, and b) fluctuations of the latter to be added in suitable form to the classical motion. The strategy will be the following. Firstly, we perform a usual ("virtual") variation on a classical Lagrangian. Then, we shall put the resulting continuity equation into a form that allows us to glean general expressions for the fluctuating energy and momentum components, respectively. Assumed to now represent "real" variations, the latter will then be re-introduced as additional elements into a correspondingly modified Lagrangian. Renewed variation of the thus obtained Lagrangian will then provide the Schrödinger equation.

It was established in 1905 by Albert Einstein, through his correct description of the "photoelectric effect", i.e., that light "particles" ("photons") of frequency $\omega$ exchange with their environment only discretized chunks of energy such that the energy of a "particle" can be written as

$$E = \hbar\omega, \tag{2.1}$$

with $\hbar$ being Planck's constant $h$ divided by $2\pi$. In 1927, then, Louis de Broglie showed that any "particle" has also wave-like attributes such that its momentum $\mathbf{p}$ can be written in terms of the associated wave-number $\mathbf{k}$,

$$\mathbf{p} = \hbar\mathbf{k}. \tag{2.2}$$

In what follows, I shall use only expressions (2.1), (2.2), and the drawing on the essentials of chapter 1 of this paper, i.e., the orthogonality of wave- and particle-related vectors in classical mechanics, to derive quantum mechanics by showing how this orthogonality is generally lost, and under which circumstances this is so.

As is usual in the standard Hamilton-Jacobi formulation of classical mechanics, we can start with a formulation of the action integral for a single free particle in an $n$-dimensional configuration space. It can be formulated with the Lagrangian $L$ (and with the introduction of some external potential $V$) as

$$A = \int L d^n x dt = \int P \left\{ \frac{\partial S}{\partial t} + \frac{1}{2m} \nabla S \cdot \nabla S + V \right\} d^n x dt. \tag{2.3}$$

Here the action function $S(\mathbf{x}, t)$ is related to the particle velocity $\mathbf{v}$ via

$$\mathbf{v} = \frac{1}{m}\nabla S, \tag{2.4}$$

with the probability density $P(\mathbf{x},t)$ that a particle is found in a given volume of configuration space being normalized such that

$$\int P d^n x = 1. \tag{2.5}$$

Let us now perform a "virtual variation" of the Lagrangian in equation (2.3) and consider its result. Upon fixed end-point variation in $S$, i.e., with $\delta S = 0$ at the boundaries, one has

$$\frac{\partial L}{\partial S} - \frac{\partial}{\partial x_i}\left\{\frac{\partial L}{\partial\left(\frac{\partial S}{\partial x_i}\right)}\right\} = 0, \tag{2.6}$$

where the index $i$ runs over the time and the three spatial components, respectively. Thus one obtains the "continuity equation" for some probability distribution $P$, i.e.,

$$\frac{\partial P}{\partial t} = -\nabla\cdot(\mathbf{v}P). \tag{2.7}$$

For the time being, let us at first restrict ourselves to considering the stationary state of constant flow only, such that the l.h.s. of (2.7) is equal to zero. Then we have after division by $P$ that

$$\nabla\cdot\mathbf{v} = -\frac{\nabla P}{P}\cdot\mathbf{v}. \tag{2.8}$$

We observe that the classical, so-called "Hamiltonian flow" (i.e., of incompressible fluids) given by

$$\nabla\cdot\mathbf{v} = 0 \tag{2.9}$$

is only obtained if the r.h.s. of (2.8) vanishes, too, i.e.,

$$\left(\frac{\nabla P}{P}\right)\cdot\mathbf{v} = 0. \tag{2.10}$$

Thus, unless trivially $\nabla P = 0$, Hamiltonian flow can also be characterized by an orthogonality of two vectors, i.e., the velocity $\mathbf{v} = \nabla S/m$ and the vector

$$\frac{\nabla P}{P} =: \text{const}\cdot\mathbf{u} = \text{const}\cdot\frac{\nabla(\delta S)}{m}, \tag{2.11}$$

which can also be set as proportional to a velocity $\mathbf{u}$. In fact, the totality of all vectors $\mathbf{u} = \delta(\nabla S/m) = \nabla(\delta S)/m$ orthogonal to $\mathbf{v}$ represents the velocity field of the spherical wave fronts mentioned in Chapter 1, which can be considered to permanently emanate from the particle as Huygens waves. Necessarily, because of the balancing spherical symmetry, the corresponding momentum $m\mathbf{u}$ in Hamiltonian flow does not generate a change in the particle's momentum $m\mathbf{v}$. (Fig. 2, left side)

In general, however, equation (2.8) is an expression for the non-conservation of momentum $\mathbf{p} = m\mathbf{v}$. Upon the performed "virtual" variation of the action function $S$, we can now work out an expression for a "real" variation $\delta S$, which does exist as a physical possibility once the "classical" orthogonality condition is discarded, and which involves some additional momentum $\delta \mathbf{p}$. This is due to the fact that we have to do with a probability field $P$, which in turn refers to stochastic processes such as a possible momentum fluctuation $\delta \mathbf{p}$.

Now let us return to the full equation (2.7). Multiplication by $dt$ and division by $P$ provides that

$$\frac{\partial}{\partial t} \ln P \, dt = -\left( \frac{\nabla P}{P} \cdot \mathbf{v} + \nabla \cdot \mathbf{v} \right) dt. \tag{2.12}$$

The l.h.s. of (2.12) is an expression for the relative temporal variation $\delta_t P/P$, which, if vanishing, is identical to an extremal principle, i.e., that the relative change of $P$ be extremal within the time span $dt$. In this case, also the r.h.s. must express an extremal principle, which can easily be identified with a "minimal action principle". This time, however, as opposed to equation (1.2), we are not dealing with the constant action function itself, but with a variation $\delta S = \text{constant}$ of it. Thus, for time intervals when the l.h.s. of (2.12) vanishes, it holds that

$$d(\delta S) = -\alpha \hbar \left\{ \frac{\nabla P}{P} \cdot d\mathbf{x} + (\nabla \cdot \mathbf{v}) \, dt \right\} = 0. \tag{2.13}$$

In (2.13), we decompose the proportionality factor into a constant $\alpha$ to be either derived or postulated, and into Planck's constant, respectively. The latter appears naturally, as any change in $\delta S$ must be quantized in the same way as energy and momentum, respectively. As the first term on the r.h.s. of (2.13) already via (2.11) indicates a fluctuating momentum $\delta \mathbf{p} = -\alpha \hbar \frac{\nabla P}{P}$, the second term must be proportional to some expression $-\frac{\partial}{\partial t}(\delta S) = \delta E$, i.e., involving some scalar expression for a fluctuating energy term. In fact, as is shown in the Appendix, the latter can be identified with the "zero-point energy" $\delta E =: E_0 = 3\frac{\hbar \omega}{2}$, thereby providing that

$$(\nabla \cdot \mathbf{v}) dt = -2 \frac{\partial}{\partial t}\left( \frac{\delta S}{\hbar} \right) dt. \tag{2.14}$$

Comparison of the coefficients then provides that the constant in (2.13) is given by

$\alpha = 1/2$, such that with (2.11)

$$\mathbf{u} = \frac{\nabla(\delta S)}{m} = -\frac{\hbar}{2m}\frac{\nabla P}{P}. \tag{2.15}$$

(Note that for our derivation of the Schrödinger equation, the main purpose of the calculation in the Appendix is given by a determination of the constant in equation (2.13), and that explicit reference to the zero-point energy is not needed for said derivation. For the latter, it is sufficient to postulate the size of the constant, or the straightforward ansatz (2.14), respectively. The appearance of Planck's constant is particularly straightforward, as any fluctuations in energy or momentum must follow the same quantizations in terms of $\hbar$ as in equations (2.1) and (2.2), respectively.)

Moreover, we can now put equation (2.12) into the form

$$\frac{\partial}{\partial t}\ln P\,dt = \frac{2}{\hbar}d(\delta S) = \frac{2}{\hbar}(\delta\mathbf{p}\,d\mathbf{x} - \delta E\,dt). \tag{2.16}$$

Just as in equation (1.2) for wave-planes with $S = \text{constant}$, one can thus for the fluctuating terms consider additional wave-planes with $\delta S = \text{constant}$, i.e., for the time intervals where $\delta_t P/P = 0$:

$$d(\delta S) = \nabla(\delta S)\cdot d\mathbf{x} + \left(\frac{\partial(\delta S)}{\partial t}\right)dt = |\nabla(\delta S)|\mathbf{n}\cdot d\mathbf{x} + \left(\frac{\partial(\delta S)}{\partial t}\right)dt = 0, \tag{2.17}$$

where

$$\mathbf{n} = \nabla(\delta S)/|\nabla(\delta S)| \tag{2.18}$$

is a unit vector perpendicular to the surfaces of $\delta S = \text{constant}$ at point $\mathbf{x}$. Thus, with equations (2.12) and (2.16) we can write out explicitly

$$d(\delta S) = -\frac{\hbar}{2}\left(\frac{\nabla P}{P}\cdot\mathbf{v} + \nabla\cdot\mathbf{v}\right)dt. \tag{2.19}$$

As it follows from (2.16) that

$$d\ln P = \frac{\partial}{\partial t}\ln P\,dt + \mathbf{v}\cdot\nabla\ln P\,dt = \frac{\partial}{\partial t}\ln P\,dt + \nabla\ln P\,d\mathbf{x} = (\nabla\cdot\mathbf{v})dt, \tag{2.20}$$

one arrives at the familiar relation rendering $P = \text{const}$ for $\nabla\cdot\mathbf{v} = 0$, i.e.,

$$P = P_0 e^{-\int(\nabla\cdot\mathbf{v})dt}. \tag{2.21}$$

However, the explicit form (2.13), or (2.19), respectively, provides us with an expression for the envisaged momentum fluctuation as given by

$$\delta \mathbf{p} = \nabla(\delta S) = -\frac{\hbar}{2}\left|\frac{\nabla P}{P}\right|\mathbf{n}. \qquad (2.22)$$

Moreover, if we make use of the result of the Appendix regarding the introduction of the "zero-point energy", the corresponding energy fluctuation term is given by

$$\delta E = E_0 = \frac{\hbar}{2}(\nabla \cdot \mathbf{v}) = 3\frac{\hbar\omega}{2}. \qquad (2.23)$$

In slightly different terms, i.e., by using the relevant wave-numbers, we arrive at the following physical picture. Firstly, we remind ourselves of the presence of Hamilton-Jacobi waves surrounding the particle, and that the intensity of a (classical) wave field with amplitude $R$ can be identified with the probability field via

$$P(\mathbf{x},t) = R^2(\mathbf{x},t). \qquad (2.24)$$

Then, in classical Hamiltonian flow, a particle has momentum $\mathbf{p} = \hbar\mathbf{k}$, and from its actual position there are waves emitted where the wave-numbers are simply

$$\mathbf{k}_u = \frac{\delta \mathbf{p}}{\hbar} = -\frac{1}{2}\left(\frac{\nabla P}{P}\right) = -\frac{\nabla R}{R} = -\left|\frac{\nabla R}{R}\right|\mathbf{n}, \qquad (2.25)$$

i.e., these waves are emitted in such a way that the wave-surfaces with action function $S = \text{constant}$ and the particle trajectories along $\nabla S$ are always orthogonal. (cf. Fig. 2, left)

However, the more interesting situation is the one where this orthogonality is disturbed. We are then facing a *contextual* situation, where not only the waves originating from the particle position are relevant, but where said waves interfere or are in some other way related to other waves from the surrounding environment. In other words, we then deal with *constraints* on our system, which will have an effect on the probability distribution $P$ across the whole (configuration space) volume which is relevant for the system's dynamics. And, as $P$ will be affected, so will be our wave number $\mathbf{k}_u$, according to equation (2.25). The non-orthogonality of $\mathbf{k}_u$ and $\mathbf{k}$ then implies that the momentum $\mathbf{p} = \hbar\mathbf{k}$ is not a conserved quantity any more. Rather, momentum fluctuations have to be taken into account, which means that the direction $\mathbf{n}$ of the vector (2.25) may change spontaneously and uncontrollably. In other words, whenever $\mathbf{k}_u$ is not orthogonal to $\mathbf{k}$, but instead allowed to fluctuate randomly (i.e., either due to "internal" spontaneous action, or because of a contextual shift in the $P$-field), the momentum $\mathbf{k}$ itself will also undergo a fluctuation, a fact which will be shown below to lead from "classical" to "quantum" motion. (cf. Fig. 2, right)

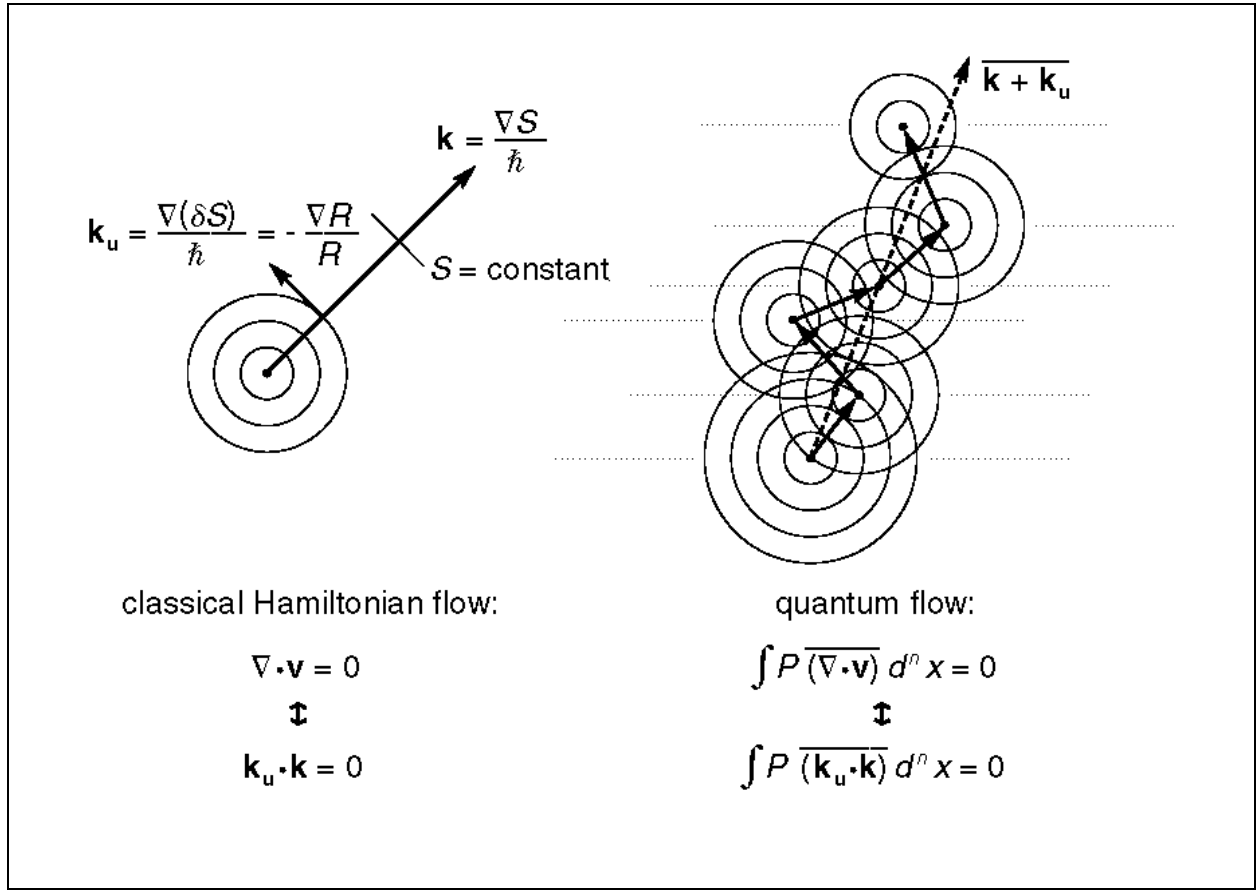

Fig.2: Schematic distinction of classical Hamiltonian flow (left) and quantum flow (right), with the circles indicating the propagation of spherical Hamilton-Jacobi wave surfaces at arbitrary distances. The essential difference is given by a vanishing divergence of the velocity of the probability current in the former, whereas the latter results from a less stringent requirement, i.e., that only the average over fluctuations and positions of the average divergence be identical to zero. The dotted lines in the figure on the right indicate symbolically that the waves pictured represent only the local surroundings of a generally extending probability field, thus illustrating that the fluctuations shown are to be seen in the context of the whole embedding environment.

We are thus lead to conclude that $\mathbf{p} = \nabla S$ can only be an average momentum which is subject to momentum fluctuations $\delta \mathbf{p}$ around $\nabla S$. Thus, the physical momentum is given by

$$\mathbf{p} := \nabla S + \delta \mathbf{p}. \tag{2.26}$$

Denoting the *average* of a variable like that of the momentum fluctuation with a bar, $\overline{\delta \mathbf{p}}$, we shall make the natural assumption that the average over fluctuations and position of it vanishes identically, i.e.,

$$\int P \overline{\delta \mathbf{p}} d^n x = 0. \tag{2.27}$$

In other words, the fluctuations are considered to be unbiased.

We are now ready to continue our intended program, i.e., to re-insert the additional fluctuating quantities into a correspondingly modified Lagrangian. In general terms, we now have

$$A := \int P \left\{ \overline{\frac{\partial S}{\partial t}} + \frac{\overline{\mathbf{p}} \cdot \overline{\mathbf{p}}}{2m} + V \right\} d^n x\, dt, \qquad (2.28)$$

with $\overline{\dfrac{\partial S}{\partial t}} = \dfrac{\partial S}{\partial t} + \overline{\dfrac{\partial (\delta S)}{\partial t}}$ and $\overline{\mathbf{p}} = \overline{\nabla S + \delta \mathbf{p}}$. With equation (2.23), we can write

$$A = \int P \left\{ \frac{\partial S}{\partial t} - \frac{\hbar}{2} \overline{(\nabla \cdot \mathbf{v})} + \frac{1}{2m} \overline{\nabla S + \delta \mathbf{p}} \cdot \overline{\nabla S + \delta \mathbf{p}} + V \right\} d^n x\, dt. \qquad (2.29)$$

Further, comparing with classical Hamiltonian flow, the respective relation $\nabla \cdot \mathbf{v} = 0$ will have to be relaxed in order to allow for the fluctuations (cf. Fig. 2). Thus, with the natural demand that the average of the velocity divergence be unbiased, we obtain

$$\int P \overline{(\nabla \cdot \mathbf{v})}\, d^n x = 0. \qquad (2.30)$$

It then follows, along with the fact that the integral $\int P \overline{\left(\dfrac{\partial}{\partial t} \ln P\right)} d^n x$ must definitely vanish when considering the (unbiased) average relative changes in time of $P$, that with (2.12) and (2.30) we obtain

$$\int P \overline{(\nabla S \cdot \delta \mathbf{p})}\, d^n x = \hbar^2 \int P \overline{(\mathbf{k} \cdot \mathbf{k}_u)}\, d^n x = 0. \qquad (2.31)$$

Thus, just as in a similar argument in [6], the average momentum fluctuations $\overline{\delta \mathbf{p}}$ are linearly uncorrelated with the average momentum $\overline{\nabla S}$. (Or, *vice versa*, we could *derive* equation (2.30) by simply postulating (2.31).)

Summarizing, the second term of (2.29) vanishes identically, and we can formulate the new action integral of our particle as

$$A := \int P \left\{ \frac{\partial S}{\partial t} + \frac{\overline{\mathbf{p}} \cdot \overline{\mathbf{p}}}{2m} + V \right\} d^n x\, dt. \qquad (2.32)$$

With condition (2.31), this provides

$$A = \int P \left\{ \frac{\partial S}{\partial t} + \frac{1}{2m} \nabla S \cdot \nabla S + V \right\} d^n x\, dt + \frac{1}{2m} \int \left(\Delta \overline{\mathbf{p}}\right)^2 dt, \qquad (2.33)$$

where $\Delta \overline{\mathbf{p}}$ turns out to equal the average rms momentum fluctuation, i.e.,

$$\left(\Delta \overline{\mathbf{p}}\right)^2 = \int P \overline{\delta \mathbf{p} \cdot \delta \mathbf{p}} \, d^n x = \int P |\delta \mathbf{p}|^2 d^n x. \tag{2.34}$$

Now, although we cannot know the direction of the assumed fluctuation, $\mathbf{n} = \delta \mathbf{p}/|\delta \mathbf{p}|$, which is subject to some (yet unknown) sub-quantum stochastic process, we actually know the general expression for its size $|\delta \mathbf{p}|$ from equation (2.22), such that

$$|\delta \mathbf{p}| = \hbar |\mathbf{k}_u| = \frac{\hbar}{2} \left| \frac{\nabla P}{P} \right|. \tag{2.35}$$

Inserting (2.35) into (2.34), we get

$$\left(\Delta \overline{\mathbf{p}}\right)^2 = \int P \left( \frac{\hbar}{2} \left| \frac{\nabla P}{P} \right| \right)^2 d^n x. \tag{2.36}$$

Thus, combining equations (2.33) and (2.36), we obtain the new (and final) action integral for our "contextual" particles as

$$A := \int P \left\{ \frac{\partial S}{\partial t} + \frac{(\nabla S)^2}{2m} + \frac{\hbar^2}{8m} \left( \frac{\nabla P}{P} \right)^2 + V \right\} d^n x \, dt. \tag{2.37}$$

Now, upon renewed fixed end-point variation in $S$ nothing changes, i.e., we again obtain the continuity equation (2.7). However, performing now the fixed end-point variation in $P$, i.e.,

$$\frac{\partial L}{\partial P} - \frac{\partial}{\partial x_i} \left\{ \frac{\partial L}{\partial \left( \frac{\partial P}{\partial x_i} \right)} \right\} = 0, \tag{2.38}$$

one obtains the so-called Hamilton-Jacobi-Bohm equation, i.e.,

$$\frac{\partial S}{\partial t} + \frac{(\nabla S)^2}{2m} + V + \frac{\hbar^2}{4m} \left[ \frac{1}{2} \left( \frac{\nabla P}{P} \right)^2 - \frac{\nabla^2 P}{P} \right] = 0. \tag{2.39}$$

The last term on the l.h.s. of equation (2.39) in the usual deBroglie-Bohm interpretation is called the "quantum potential" [7,8], but I want to stress that this "non-classical" term in the context presented here is due to a kinetic energy term (viz., equations (2.36) and (2.37)) rather than a potential energy, as was similarly argued previously by Harvey [9].

Still, as is well known [7,8], the equations (2.7) and (2.39), together with the introduction of the complex-numbered "wave function"

$$\psi = \sqrt{P} \exp\{-iS/\hbar\}, \tag{2.40}$$

can be condensed into a single equation, i.e., the Schrödinger equation

$$i\hbar \frac{\partial \psi}{\partial t} = \left(-\frac{\hbar^2}{2m}\nabla^2 + V\right)\psi. \tag{2.41}$$

Extension to a many-particle system is straightforwardly achieved by starting the same procedure with a correspondingly altered Lagrangian in (2.37), which then ultimately provides the usual many-particle Schrödinger equation.

Moreover, we are now ready to consider also a relativistic version of the foregoing derivation, i.e., for simplicity studying the case of spinless particles. With the introduction of four-vectors, equations (2.1) and (2.2) can be written in the compact form

$$p^\mu = \left(\frac{E}{c}, \mathbf{p}\right) = \hbar k^\mu = \hbar\left(\frac{\omega}{c}, \mathbf{k}\right), \tag{2.42}$$

thus introducing the energy $E$ as the zero-component of the four-momentum $p^\mu$. Then, with the four-vector notation $dx^\mu := (cdt, d\mathbf{x})$, the usual Einstein sum convention, and in accordance with relativistic kinematics, we can formulate a relativistic Lagrangian $L$ as

$$A = \int L d^n x dt = \int P\left\{\frac{1}{m}\partial_\mu S \partial^\mu S + \frac{\partial S}{\partial t}\right\} d^n x dt = \int P\left\{\frac{1}{m}p_\mu p^\mu + \frac{\partial S}{\partial t}\right\} d^n x dt. \tag{2.43}$$

Here the action function $S(\mathbf{x},t)$ is related to the particle velocity $v^\mu = p^\mu/m$ via

$$v^\mu = \frac{1}{m}\partial^\mu S, \tag{2.44}$$

with the usual four-derivative $\partial^\mu := \left(\frac{1}{c}\frac{\partial}{\partial t}, \nabla\right)$.

Let us now perform again a "virtual variation" of the Lagrangian (i.e., this time in equation (2.43) ) and consider its result. Upon fixed end-point variation, i.e., $\delta S = 0$ at the boundaries, one has

$$\frac{\partial L}{\partial S} - \partial_\mu \left\{\frac{\partial L}{\partial(\partial_\mu S)}\right\} = 0, \tag{2.45}$$

and thus one obtains the relativistically invariant "continuity equation" for the conservation of the probability current $J^\mu = P\partial^\mu S$ of some probability distribution $P$, i.e.,

$$\partial_\mu J^\mu = \partial_\mu \left[P\partial^\mu S\right] = \partial_\mu P \partial^\mu S + P\partial_\mu \partial^\mu S = 0. \tag{2.46}$$

Inserting the expression for the four-momentum

$$p^\mu = m v^\mu = \partial^\mu S \tag{2.47}$$

into equation (2.46) then provides that the four-gradient of $p^\mu$ is given by

$$\partial_\mu p^\mu = -\frac{\partial_\mu P}{P} p^\mu. \tag{2.48}$$

Thus, very similarly to the non-relativistic case, the conservation of the probability current $J^\mu$ entails the conservation of energy-momentum, $\partial_\mu p^\mu = 0$, *only* when the scalar product on the r.h.s. of (2.48) vanishes, i.e., when the two four-vectors are orthogonal, i.e., $\frac{\partial_\mu P}{P} p^\mu = 0$. Moreover, if for conservative systems one writes out (2.48) without the use of four-vectors, one arrives again at equation (2.12)! Consequently, also the further procedures follow the corresponding one of the non-relativistic case very similarly.
In fact, the relativistic four-momentum fluctuation term now simply becomes

$$\left|\delta p^\mu\right| = \left|\hbar k_u^\mu\right| = \left|\frac{\hbar}{2}\frac{\partial^\mu P}{P}\right|, \tag{2.49}$$

and the action integral is changed into

$$A = \int L d^n x dt = \int P\left\{\frac{1}{m}\overline{\left(p_\mu + \delta p_\mu\right)\left(p^\mu + \delta p^\mu\right)} + \frac{\partial S}{\partial t}\right\} d^n x dt. \tag{2.50}$$

Using (2.50) and the analogy to the non-relativistic case that

$$\int P\overline{\left(\partial_\mu p^\mu\right)} d^n x = -\int P\left(\overline{\frac{\partial_\mu P}{P} p^\mu}\right) d^n x = 0, \tag{2.51}$$

this leads again to a correspondingly modified action integral, i.e.,

$$A = \int L d^n x dt = \int P\left\{\frac{1}{m}\partial_\mu S \partial^\mu S + \frac{\hbar^2}{4m}\frac{\partial_u P}{P}\frac{\partial^\mu P}{P} + \frac{\partial S}{\partial t}\right\} d^n x dt, \tag{2.52}$$

finally providing via the Euler-Lagrange equations (and with the usual quabla operator $\Box := \partial_\mu \partial^\mu$)

$$\partial_\mu S \partial^\mu S := M^2 c^2 = m^2 c^2 + \hbar^2 \frac{\Box \sqrt{P}}{\sqrt{P}}. \quad (2.53)$$

Again, as is well known, equations (2.46) and (2.53) can be written in compact form by using the "wave function" $\Psi = \sqrt{P} \exp(-iS/\hbar)$ to obtain the usual Klein-Gordon equation

$$\left( \Box + \frac{m^2 c^2}{\hbar^2} \right) \Psi = 0. \quad (2.54)$$

We have thus succeeded in deriving the Schrödinger- and Klein-Gordon equations from classical Lagrangians with a minimal set of additional assumptions relating to an additional momentum fluctuation (viz., equations (2.35) or (2.49), respectively) associated to each particle.

### 3. Derivation of Heisenberg's uncertainty relations

We have seen that the introduction of a momentum fluctuation field is necessary to obtain a complete description of a "particle in an environment". So, even if for the time being we assumed that our knowledge of the particle's momentum were given in one part by the exact classical momentum (i.e., with infinite precision), we must still consider the latter to be "smeared" by the presence of the momentum fluctuation $\delta \mathbf{p}$ in equation (2.26), such that the uncertainty in our knowledge of the particle's momentum $\delta p_0$ would then be given by the average rms momentum fluctuation according to equation (2.36), i.e., restricting ourselves to one dimension for simplicity,

$$\delta p_0 := \Delta \overline{p} = \sqrt{\int P \left( \frac{\hbar}{2} \frac{\nabla P}{P} \right)^2 dx}. \quad (3.1)$$

Now we recall that a classical measure of minimal position uncertainty, as derived from the "Fisher information" [10], is given by the "Fisher length" [6]

$$\delta x = \left[ \int P \left( \frac{\nabla P}{P} \right)^2 dx \right]^{-1/2}. \quad (3.2)$$

Comparing (3.1) and (3.2) immediately provides the "exact uncertainty relation" of Hall and Reginatto [6],

$$\delta x = \frac{1}{\sqrt{\int P \left( \frac{\hbar}{2} \frac{\nabla P}{P} \right)^2 dx}} \frac{\hbar}{2} = \frac{1}{\delta p_0} \frac{\hbar}{2},$$

such that

$$\delta x \delta p_0 = \frac{\hbar}{2}. \qquad (3.3)$$

This exact uncertainty relation holds only in a limiting case, however. In fact, if we now admit the general uncertainty in our knowledge of the total momentum, $\Delta p$, to come from both momentum contributions involved, i.e., according to equation (2.26),

$$\Delta p := \delta(\nabla S) + \delta p_0, \qquad (3.4)$$

we obtain that

$$\Delta p \geq \delta p_0. \qquad (3.5)$$

Moreover, according to the Cramer-Rao inequality of statistical estimation theory [11], it holds that the variance of any estimator $\Delta x$ is equal to, or larger, than the optimal variance, which is given by the Fisher length, i.e.,

$$\Delta x \geq \delta x. \qquad (3.6)$$

Therefore, combining equations (3.3), (3.5), and (3.6), one obtains Heisenberg's uncertainty relations

$$\Delta x \Delta p \geq \frac{\hbar}{2}. \qquad (3.7)$$

Thus, the uncertainty relations are physically explained by the "smearing out" of a particle's classical momentum due to the momentum fluctuation field of the surrounding environment.

Moreover, the form of equation (3.1) already hints at the recently established result [12] that the uncertainty relations are but a special consequence of the more powerful general statement that a quantum state is (nonlocally) entangled with the apparatus. Since the momentum fluctuation in (3.1) depends only on the *relative gradient* of $P$, its expression does not necessarily fall off with any distance between component parts of a probability distribution. In other words, even small relative changes may become fully effective across nonlocal distances.

### 4. Summary and Outlook

We can now consider a summary of this paper's main results, thereby putting particular emphasis on the comparison between classical Hamiltonian flow and quantum theory. One may start with the classical conservation law for a probability current in configuration space,

$$\frac{\partial P}{\partial t} = -\nabla(\mathbf{v} P), \qquad (4.1)$$

which after division by $P$ provides

$$\frac{\partial}{\partial t}\ln P = -\left(\frac{\nabla P}{P}\cdot\mathbf{v} + \nabla\cdot\mathbf{v}\right). \tag{4.2}$$

It is usually argued that Hamiltonian flow, i.e., $\nabla\cdot\mathbf{v} = 0$, follows from $P = \text{constant}$. [13] However, we have seen that for Hamiltonian flow to occur, the minimal requirement for $P$ is that only $\frac{\partial}{\partial t}\ln P = 0$ holds always, and that the orthogonality condition $\mathbf{v}\cdot\nabla\ln P = 0$ is fulfilled. This, in turn, is in perfect agreement with the physical picture of Huygens-waves emanating from the "particle", which are characterized by spherically symmetric distributions of equal-sized momentum fluctuations

$$\nabla(\delta S) = \hbar\mathbf{k}_u = -\alpha\hbar\nabla\ln P \tag{4.3}$$

(with $\alpha$ being some proportionality constant), thus obeying the above mentioned orthogonality condition and providing no net contribution to the particle's momentum $\hbar\mathbf{k}$. However, as the probability density is related to the intensity of wave-like distributions via $P(\mathbf{x},t) = R^2(\mathbf{x},t)$, where $R$ signifies the wave amplitude, we can propose that the wave-number $\mathbf{k}_u$ is simply a gradient vector measuring the relative spatial decrease of $R$ away from the original particle position, i.e.,

$$\mathbf{k}_u = -\frac{\nabla R}{R}. \tag{4.4}$$

This fixes the constant $\alpha$ in (4.3) as $\alpha = 1/2$. (Thus, in this summary, we do not need to refer to the "zero-point energy argument" of the Appendix.)

As already mentioned, the case of classical Hamiltonian flow is described by equation (4.2), with each term vanishing identically. However, it can also be shown straightforwardly, that the essential difference to quantum theory is given simply by taking the average of each term of (4.2), and integrating the resulting expressions over $P$ and the configuration space. Then, from (4.2) one gets (with bars denoting averages):

$$\int P \overline{\frac{\partial}{\partial t}\ln P}d^n x = -\int P\overline{\left(\frac{\nabla P}{P}\cdot\mathbf{v}\right)}d^n x - \int P\overline{(\nabla\cdot\mathbf{v})}d^n x. \tag{4.5}$$

As the l.h.s. naturally vanishes identically, because only *unbiased averages* over the average temporal behavior of $P$ are plausible, one can also naturally demand that the expressions $\nabla\ln P$ and $\mathbf{v}$, respectively, are linearly independent, thus implying that also both terms on the r.h.s. of equation (4.5) vanish identically. In other words: *whereas classical Hamiltonian flow is described by equation* (4.2), *with each term vanishing identically, quantum theory follows from equation* (4.5), *with all its terms equal to zero.*

In fact, with the choice of (4.4), and thus $\alpha = 1/2$ in equation (4.3), one can rewrite (4.2) as

$$\frac{\partial}{\partial t}\ln P dt = \frac{2}{\hbar}d(\delta S) = \frac{2}{\hbar}(\delta\mathbf{p}d\mathbf{x} - \delta E dt), \tag{4.6}$$

which thus provides us with the looked-for expressions for the fluctuating momentum and energy terms, respectively:

$$\delta\mathbf{p} = \nabla(\delta S) = -\frac{\hbar}{2}\frac{\nabla P}{P}, \quad \delta E = \frac{\hbar}{2}(\nabla \cdot \mathbf{v}). \tag{4.7}$$

Using (4.7) to re-define the *average* quantities of momentum and energy, respectively, one proceeds to insert the latter into a thus modified classical Lagrangian. Then, with the established conditions for the averages as mentioned w.r.t. equation (4.5), renewed variations via Euler-Lagrange equations provide the familiar current conservation and Hamilton-Jacobi equations which ultimately lead to Schrödinger's equation.

It is now time to make a remark on the expression (2.15) for the velocity field related to the momentum fluctuation, i.e.,

$$\mathbf{u} = \frac{\nabla(\delta S)}{m} = -\frac{\hbar}{2m}\frac{\nabla P}{P}. \tag{4.8}$$

As is well-known, this is the expression for the diffusion velocity in standard spatial Brownian motion, i.e., where $\mathbf{u} = -\tilde{\mathbf{u}}$, with $\tilde{\mathbf{u}}$ called the "osmotic velocity" in the stochastic interpretation of quantum theory. [14, 15] In fact, the decomposition of the Schrödinger equation into two real-valued ones, i.e., equations (4.1) and (2.39), can also be taken as the starting point of the stochastic theory. For, taking the gradients of said equations, one gets

$$\frac{\partial \tilde{\mathbf{u}}}{\partial t} = -D\nabla^2 \mathbf{v} - \nabla(\tilde{\mathbf{u}} \cdot \mathbf{v}) \tag{4.9}$$

and

$$\frac{\partial \mathbf{v}}{\partial t} = \mathbf{a} - (\mathbf{v}\cdot\nabla)\mathbf{v} + (\tilde{\mathbf{u}}\cdot\nabla)\tilde{\mathbf{u}} + D\nabla^2\tilde{\mathbf{u}}, \tag{4.10}$$

with the "diffusion coefficient" $D := \frac{\hbar}{2m}$, and the classical acceleration $\mathbf{a} = -\frac{\nabla V}{m}$. Equations (4.9) and (4.10), together with (4.1), represent nothing but the coupled differential equations for Brownian motion of a particle suspended in some "medium". One look at the quantum part of Fig. 2 shows exactly why this comes as no big surprise: The "spontaneous" fluctuations we have dealt with in this paper might very well, at least in part, stem from the dynamics of some sub-quantum medium, or "aether", which means that the motion of a particle can also be seen as a diffusion process through this medium.

Finally, we also see that the momentum fluctuations can either be induced via diffusion processes from the (generally non-local) $P$-field, or via local spontaneous action (like, e.g., the collision of the "particle" with one of the aether's assumed sub-quantum constituents). In other words, a quantum "particle" can both be "guided" by the surrounding context, or it can alter the latter via said spontaneous fluctuations. This is also why one can speak [16,17] of a mutual dependence, or "circular causality", between the particle and its environment.

**Appendix: Determination of the constant $\alpha$ in equation** (2.13)

The following presents one (optional) way to determine $\alpha$. (Another one is given in the summary of the paper.) Let us start with an experimental finding from the year 1924 [18]: It was then found that to each energy $E = \hbar\omega$ of a particle, one must associate (in three spatial dimensions) an additional term of a "zero-point energy"

$$E_0 = 3\frac{\hbar\omega}{2}, \tag{5.1}$$

which refers to a fluctuating field in the vacuum.
Further, we note that the second summand in equation (2.12) can be written as

$$(\nabla \cdot \mathbf{v})dt = \left\{ \frac{\partial}{\partial x} \cdot \frac{1}{dt} dx + \frac{\partial}{\partial y} \cdot \frac{1}{dt} dy + \frac{\partial}{\partial z} \cdot \frac{1}{dt} dz \right\} dt. \tag{5.2}$$

If we now observe that with equation (2.1) the quantum system has a "characteristic frequency" $\omega$, this frequency also determines the maximal temporal resolution of the system (i.e., such that time spans shorter than one undulatory period of a wave, $\tau = 1/\omega$, are not operational), we obtain a "coarse graining" of the time axis, with its inverse elementary unit becoming

$$\frac{1}{dt} \to \frac{1}{\tau} = \omega. \tag{5.3}$$

Thus, $1/dt$ is substituted by $\omega$, and we get from (5.2) that

$$(\nabla \cdot \mathbf{v})dt = 3\omega dt. \tag{5.4}$$

Comparison with (5.1) then provides that

$$-\frac{\partial(\delta S)}{\partial t} = \delta E = \frac{3}{2}\hbar\omega = \frac{1}{2}\hbar(\nabla \cdot \mathbf{v}), \tag{5.5}$$

and we thus obtain that in equation (2.13), $\alpha = \frac{1}{2}$.